\documentclass[12pt,onecolumn,showpacs,preprintnumbers]{revtex4}
\usepackage{graphicx}
\begin{document}

\title{Cluster formation in the Fermi system with long-range interaction}
\author{K.V. Grigorishin}
\email{konst@orta.net.ua}
\author{B.I. Lev}
\email{blev@i.kiev.ua}

\address{Bogolyubov Institute for Theoretical Physics of the
Ukrainian National Academy of Sciences, 14-b Metrolohichna str.
Kiev-03680, Ukraine.}

\pacs{ 05.30-d, 05.70.Fh}

\begin{abstract}
Based on statistical approach we described possible formation of
spatially inhomogeneous distribution in the system of interacting
Fermi particles by long-rage forces, and we demonstrated
nonperturbative calculation of the partition function in this
case. It was shown, that particles interacting with an attractive
$1/r$ potential form clusters. Cluster is equilibrium structure,
if we suppose that average energy of interaction of two particles
is much less than their average kinetic energy $kT$. The analogy
between  self-gravitation gas and plasma was shown. The dynamics
of cluster formation was considered with help hydrodynamical and
statistical approaches, and time of relaxation to equilibrium
state was found.
\end{abstract}

\maketitle

\section{Introduction.}

The formation of spatially inhomogeneous distribution of
interacting particles is a typical problem in condensed matter
physics. The types of spatial structures and conditions for their
formation are determined by type of interaction. A cluster is one
from forms of spatial inhomogeneity. For gas with attraction of
Coulomb type between particles (self-acting system) we can't
calculate the virial coefficients (we cannot do it for interaction
$1/r^{n}$ if $n\leq 3$ \cite{rue}). Statistical description of the
such system developed in Refs. \cite{bel,bel1,zhu,lev,kras} is
based on the application of the apparatus of quantum theory of
field \cite{col,raj,lip,edw,sam,mag,veg1} where it was shown that
gas with interacting particles is equivalent to two scalar fields
corresponding attraction and repulsion accordingly with an
exponential self-action. The partition function can be
representation in terms of a functional integral over these
auxiliary fields. The extremal conditions for this functional are
nonlinear equations. The spatial distribution function, which
describes the cluster, is soliton solution of the equations. This
approach gives the possibility to find the spatial distribution of
particles, to calculate the cluster's size and to determine the
temperature of phase transition in to the state under
consideration by nonperturbative calculation.

In the papers \cite{gri1,gri2}, this approach has been applied to
investigation of the system with long-range interacting Bose
particles. It has been shown that spatial distribution of such
system is inhomogeneous and has view of finite-size cluster. The
radius of cluster, conditions and dynamics of its formation have
been found both for undegenerated states and for Bose-Einstein
condensation regime. In this article, based on statistical
approach \cite{bel,bel1,zhu,lev}, we demonstrate a nonperturbative
calculation of the partition function, we solve the system of
Fermi particles interacting with an attractive $1/r$ potential,
that is we obtain free energy for the such system. We obtain
expression for equilibrium radius of cluster and we described
dynamics of the cluster formation. However, we can find exact
results in Boltsman limit with first quantum corrections. For all
temperatures we propose the approach which gives the possibility
to find radius of the cluster approximately or to evaluate of one.

So far as the cluster's radius we can understand as size of
spatial inhomogeneity of a system, then it is present interest to
investigate the general properties of spatial distribution
function of the systems with Coulomb interacting particles -
above-mentioned gas of gravitating particles and the system of
repulsing particles. This can be done in statistical and
hydrodynamical approaches. Using hydrodynamical approaches we can
find the times of relaxation of the systems with Coulomb
interacting particles.

\section{Statistical approach to the system of interacting particles.}

Let's consider an interacting particles' system being in such
conditions when, on the one hand, wave's thermal length of a
particle can be larger than average distance between them, so that
it is necessary to take into account type of the statistic, but,
on the other hand, this length is by far smaller than average
scattering length, that lets to describe interaction classically
disregarding dynamical quantum correlations. The Hamiltonian of
such a system \cite{bel,zhu,mag,bax} is

\begin{equation}\label{1.1}
    H(n)=\sum_{s}\varepsilon_{s}n_{s}-\frac{1}{2}\sum_{ss'}W_{ss'}n_{s}n_{s'}
    +\frac{1}{2}\sum_{ss'}U_{ss'}n_{s}n_{s'}\texttt{,}
\end{equation}
where $\varepsilon_{s}$ is the additive part of particle energy in
the state $s$ (for example kinetic energy or energy in an external
field), $W_{ss'}$ and $U_{ss'}$ are the absolute values of the
attraction and repulsion energies of particles in the states $s$
and $s'$ respectively. The macroscopic state of the system is
determined by occupations numbers $n_{s}$. The subscript $s$
corresponds to variables that describe an individual particle
state.

In Refs.\cite{bel,bel1,zhu,lev} in order to investigate
thermodynamical properties of the interacting particles' system a
Hubbard-Stratonovich \cite{hub,str,don,kam} representation has
been used for the partition function in the terms of grand
partition function
\begin{equation}\label{1.2}
    Z_{n}=\frac{1}{2\pi i}\oint d\xi\int D\varphi \int D\psi \exp\left(-S\left(\xi,\varphi,\psi\right)\right)\texttt{,}
\end{equation}
where $S$ is the functional which we call an effective free energy
which is analogous to an action in field theory:
\begin{eqnarray}
S\left(\xi,\varphi,\psi\right)=&&\frac{1}{2\beta}\sum_{s,\acute{s}}
\left(W^{-1}_{s,s^{'}}\varphi_{s}\varphi_{s^{'}}+U^{-1}_{s,s^{'}}\psi_{s}\psi_{s^{'}}\right)\nonumber\\
&&+\delta\sum_{s}\ln\left(1-\delta\xi\exp\left(-\beta
\varepsilon_{s}+\varphi_{s}\right)\cos\psi_{s}\right)+\left(N+1\right)\ln\xi
\texttt{,}\label{1.3}
\end{eqnarray}
$\xi\equiv e^{\beta\mu}$ is activity, $\mu$ is chemical potential,
$\beta=1/kT$ is reverse temperature, $s$ and $s^{'}$ run all the
states of the system, $\varepsilon$ is kinetic energy, $N$ is
number of particles, $\delta =+1$ for Bose particles and $-1$ for
Fermi particles. The two auxiliary fields $\varphi$ and $\psi$ are
introduced corresponding to attraction and repulsion. The
partition function (\ref{1.2}) is written as a functional integral
over these fields. $W^{-1}_{s,s^{'}},U^{-1}_{s,s^{'}}$ are inverse
operators of the interaction:
$\omega^{-1}_{ss^{'}}=\delta_{ss^{'}}\hat{L}_{s^{'}}$ where
$\hat{L}_{s^{'}}$ is such an operator for which the interaction's
potential is Green function.

The integral (\ref{1.2}) is calculated by the method of
''saddle-point'' \cite{hau,isi} across the point determined by the
functional derivatives $\frac{\delta
S}{\delta\varphi}=\frac{\delta S}{\delta\psi}=0$ as
\begin{eqnarray}
 \frac{1}{\beta}\sum_{s'}W_{ss'}^{-1}\varphi_{s'}-\frac{\xi_{s}e^{\varphi_{s}}\cos\psi_{s}}{1-\delta\xi_{s}e^{\varphi_{s}}\cos\psi_{s}}=0 \label{1.4}\\
  \nonumber\\
 \frac{1}{\beta}\sum_{s'}U_{ss'}^{-1}\psi_{s'}+\frac{\xi_{s}e^{\varphi_{s}}\sin\psi_{s}}{1-\delta\xi_{s}e^{\varphi_{s}}\cos\psi_{s}}=0\label{1.5}
\end{eqnarray}
and the derivative
\begin{equation}\label{1.6}
 \frac{\partial S}{\partial\xi}=0\Rightarrow\sum_{s}\frac{\xi_{s}e^{\varphi_{s}}\cos\psi_{s}}{1-\delta\xi_{s}e^{\varphi_{s}}\cos\psi_{s}}=N+1\texttt{,}
\end{equation}
where $\xi_{s}=\xi e^{\beta\varepsilon_{s}}$. This set of
equations (\ref{1.4}-\ref{1.6}) provides a solution of the
many-particle problem in the sense that it selects the system
states whose contributions in the partition function are dominant.
The third equation is normalization condition for the particle
distribution function which is determined by the auxiliary fields.
It is obvious that, for given statistics, the distribution
function depends on the interaction nature and intensity. The
cluster formation corresponds to particle localization within a
limited space. In our treatment, the effect is reflected in the
behavior of the auxiliary fields and chemical potential.

For the screened Coulomb or Newtonian potential, the inverse
operator may be written as
\begin{equation}\label{1.3a}
  \widehat{L}=-\frac{1}{4\pi q^{2}}(\triangle-\lambda^{2})\texttt{,}
\end{equation}
where $q^{2}$ is the interaction constant, $\triangle$ is
Laplace's operator and $\lambda^{-1}$ is the screening length
\cite{hub,edw,sam,mag}. The number of realistic interactions, for
which the inverse operator can be found and general solution of
the set (\ref{1.4}-\ref{1.6}) can be obtained, is limited. That's
why we'll confine ourselves to long-range both attractive and
repulsive Coulomb potential ($\propto\frac{1}{R}$) for Fermi gas
where we'll find the conditions for the cluster formation and
cluster parameters.

In the continuum approximation, the subscript $s$ runs through a
continuum of values in the system of volume $V$. When integrating
over impulses and coordinates, we bear in mind that the unit of
cell's volume in the space of individual states is equal to
$\omega=(2\pi\hbar)^{3}$. In order to avoid unnecessary
complications we shall consider particles without spin only.

\section{Cluster formation in the system of attracting particles.}
\subsection{The equations for spatial distribution function for system with Coulomb attraction.}
\label{sec:Equations}

Now we shall consider the system of particles interacting by
long-range attraction only and we shall demonstrate a
nonperturbative calculation of the partition function. For Coulomb
attraction the inverse operator is known to be
$W^{-1}_{rr^{'}}=\frac{-1}{4\pi
q^{2}}\triangle_{r}\delta_{rr^{'}}$, where $g$ is the
interactional constant and $\triangle_{r}$ is the Laplace
operator.

Let's consider the system in the usual thermodynamical limit:
number of particles $N\rightarrow\infty$ and volume
$V\rightarrow\infty$ with fixed $N/V$. We consider the effective
free energy (\ref{1.3}) in spherical coordinates:
$\nabla\equiv\frac{\partial}{\partial R}$ and, neglecting surface
contribution, we can write the effective free energy (\ref{1.3})
for Fermi gas as
\begin{eqnarray}
S=&&\frac{1}{2\beta}\int_{0}^{V}\frac{\left(\nabla\varphi\right)^{2}}{4\pi
q^{2}}dV-\frac{1}{\omega}\int_{0}^{V} dV\int
d^{3}p\ln\left[\frac{}{}1+\xi e^{\varphi}
\exp\left(-\beta\frac{p^{2}}{2m}\right)\right]\nonumber\\
&&+N\ln\xi=\int_{0}^{V}dV\left[\frac{\left(\nabla\varphi\right)^{2}}{4r_{q}}-\frac{1}{\lambda^{3}}f_{5/2}\left(\xi
e^{\varphi}\right)\right]+N\ln\xi, \label{2.1}
\end{eqnarray}
where $r_{q}=2\pi q^{2}\beta$, $m$ is a particle mass,
$\lambda=\sqrt{\frac{\beta\hbar^{2}}{2\pi m}}$ is wave's thermal
length of a particle and
$f_{5/2}\left(\xi\right)\\=\frac{4}{\sqrt{\pi}}\int^{\infty}_{0}dxx^{2}\ln\left(1+\xi
e^{-x^{2}}\right)=\sum^{\infty}_{l=1}(-1)^{l+1}\frac{\xi^{l}}{l^{5/2}}$
is special Fermi function \cite{hau,isi}. Sense of the auxiliary
field $\varphi$ is next - a spatial distribution's function can be
expressed by $\varphi$ as
\begin{eqnarray}\label{2.2a}
\rho(R)=\int_{0}^{\infty}\frac{d^{3}p}{(2\pi\hbar)^{3}}
\frac{\xi\exp\left(-\frac{p^{2}}{2mkT}\right)e^{\varphi(R)}}{1+
\xi\exp\left(-\frac{p^{2}}{2mkT}\right)e^{\varphi(R)}},
\end{eqnarray}
where we used the underintegral expression in Eq.(\ref{1.6}).
Let's introduce the dimensionless quantity $r=R/r_{q}$ instead of
$R$. Then, the effective free energy (\ref{2.1}) (in spherical
coordinates) can be written as
\begin{eqnarray}
S=4\pi\int_{0}^{V}r^{2}dr\left[\frac{\left(\nabla\varphi\right)^{2}}{4}-\frac{r_{q}^{3}}{\lambda^{3}}f_{5/2}\left(\xi
e^{\varphi}\right)\right]+N\ln\xi. \label{2.2b}
\end{eqnarray}
The saddle-point equation is Lagrange equation for this
functional:
\begin{eqnarray}
\frac{\partial^{2}\varphi}{\partial
r^{2}}+\frac{2}{r}\frac{\partial\varphi}{\partial
r}+\frac{2r_{q}^{3}}{\lambda^{3}}\frac{\partial f_{5/2}\left(\xi
e^{\varphi}\right)}{\partial\varphi}=0. \label{2.3}
\end{eqnarray}
This equation selects the system's states (which are described by
the field configuration $\varphi(R)$ or spatial distribution
function $\rho[\varphi(R)]$) whose contributions in the partition
function are dominant. Unfortunately, this equation hasn't
analytical solution. But the problem simplifies in Boltzmann
limit.

\subsection{The solution in Boltzmann limit and dynamics of the cluster formation.}
\label{sec:BoltzmannLimit}

\begin{figure}
\centering
  \includegraphics[width=9.0cm]{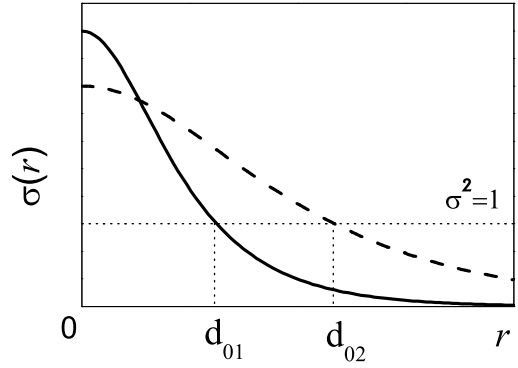}
  \caption{The spatial distribution function $\sigma(r)^{2}$ in a cluster at different
temperatures
           (schematically). The solid line corresponds to a lower
           temperature $T_{1}$, the dash line corresponds to a higher temperature $T_{2}$.
           The dot line $\sigma^{2}=1$ corresponds to mean fixed density of the system and determines
           equilibrium radii of cluster $d_{01}$ and $d_{02}$ under the above-mentioned
           thermodynamical conditions.}\label{fig1}
\end{figure}

Let's consider the limit case $\xi\rightarrow 0$ corresponding to
high temperature and small concentration, using decomposition of
the special function $f_{5/2}\left(\xi e^{\varphi}\right)$ in row
on orders of the activity $\xi$. Then the effective free
energy(\ref{2.1}) is reduced to the simpler expression
\begin{equation}\label{2.5}
  S=4\pi\int_{0}^{V}\left[\frac{1}{4}(\nabla\varphi)^{2}-
  \xi\frac{r_{q}^{3}}{\lambda^{3}}e^{\varphi} \right]r^{2}dr+N\ln\xi.
\end{equation}
That is the system with Coulomb attraction is equivalent to single
scalar field $\varphi(r)$ with an exponential self-action.
Analogous expression was obtained in Ref.\cite{veg1}, however
therm $N\ln\xi$, which fixes number of particles, is absent there.
The saddle-point equation for the functional (\ref{2.5}) is
\begin{equation}
\frac{\partial^{2}\varphi}{\partial
r^{2}}+\xi\frac{2r_{q}^{3}}{\lambda^{3}}
e^{\varphi}=0\texttt{.}\label{2.6}
\end{equation}
As it will be shown below, the term
$\frac{2}{r}\frac{\partial\varphi}{\partial r}$ can be omitted. In
order to connect the auxiliary field $\varphi$ with the density
$\rho(R)$ we have to use Eq.(\ref{2.2a}) and to pass to the
Bolzmann limit $\xi\rightarrow 0$:
\begin{equation}\label{2.6a}
\rho=\frac{\xi}{\lambda^{3}}e^{\varphi}\equiv
\frac{\xi}{\lambda^{3}}\sigma^{2},
\end{equation}
where we introduce the new variable
$\sigma=\exp\left(\varphi/2\right)$ and let's mark in
$\alpha^{2}\equiv r^{3}_{q}/\lambda^{3}$. Then we can rewrite
Eq.(\ref{2.6}) for the field $\varphi$ as equation for the density
as
\begin{equation}\label{2.6b}
    \frac{\partial^{2}\sigma}{\partial
    r^{2}}-\frac{1}{\sigma}\left(\frac{\partial\sigma}{\partial
    r}\right)^{2}+\xi\alpha^{2}\sigma^{3}=0\texttt{.}
\end{equation}
This equation has a soliton solution \cite{zhu}
\begin{equation}\label{2.7}
    \sigma=\frac{\Delta}{\sqrt{\xi}\alpha}\frac{1}{\cosh\Delta r}\texttt{,}
\end{equation}
where $\Delta$ is an integration constant. Any soliton solution
corresponds to a spatially inhomogeneous distribution of particles
- a finite-size cluster. The corresponding asymptotics are
$\sigma^{2}=1$ for $r=d$, where $d$ is the cluster size, and
$\sigma\rightarrow0$ as $r\rightarrow\infty$. This solution
describes the presence of particles in the inhomogeneous formation
of the size $d$ and the absence of particles at infinity
(Fig.\ref{fig1}). This spatial distribution is compressing to line
$\sigma=1$ or $\varphi=0$ when $T\rightarrow\infty$,
$\frac{N}{V}\rightarrow 0$, that's why the field $\varphi=0$
corresponds to spatially homogeneous distribution in the system.

Let's substitute solution (\ref{2.7}) in the effective free energy
(\ref{2.5}):
\begin{equation}
S=4\pi\int^{d}_{0}\left(\Delta^{2}-2\xi\alpha^{2}\sigma^{2}\right)r^{2}dr+N\ln\xi\texttt{.}
\label{eq:1.12}
\end{equation}
Then we'll integrate using the decomposition $1/\cosh x\approx
1-x^{2}/2$ in power series of $x\equiv\Delta d<<1$:
\begin{equation}
S=-V\frac{\Delta^{2}}{\alpha^{2}\lambda^{3}}+N\ln\xi\texttt{.}
\label{2.8}
\end{equation}
The integration constant $\Delta^{2}$ is being found from the
asymptotics $\sigma^{2}(d)=1$:
$\Delta^{2}\approx\xi\alpha^{2}+\xi^{2}d^{2}\alpha^{4}$. Thus, we
have the result:
\begin{equation}
S=-\frac{V}{\lambda^{3}}\xi+N\ln\xi-\frac{V}{\lambda^{3}}\xi^{2}
d^{2}\alpha^{2}\texttt{.} \label{2.9}
\end{equation}
Assuming that the average energy of interaction of two particles
is less than the average kinetic energy $\sim kT$ of a particle
$\frac{r^{3}_{q}N}{V}\ll 1$, we can find the activity $\xi$ from
the equation $\frac{\partial S}{\partial\xi}=0$ as
\begin{equation}
\xi\approx\frac{\lambda^{3}N}{V}-\frac{2d^{2}\alpha^{2}\lambda^{6}N^{2}}{V^{2}}\equiv
\xi_{0}+\xi_{q}\texttt{, }|\xi_{q}|\ll\xi_{0}\texttt{,}
\label{2.10}
\end{equation}
where $\xi_{0}$ and $\xi_{q}$ are activities of ideal gas and
first correction for the interaction accordingly. Then integrating
Eq.(\ref{1.2}) with the effective free energy (\ref{2.9}) on
saddle-point (\ref{2.10}) we obtain the partition function as
\begin{eqnarray}
&&Z_{N}=Z^{0}_{N}\times\exp\left[\frac{V}{\lambda^{3}}\xi_{q}-N
\ln\left(1+\frac{\xi_{q}}{\xi_{0}}\right)
+\frac{V}{\lambda^{3}}\xi^{2} d^{2}\alpha^{2}\right]\texttt{,}
\label{2.10a}
\end{eqnarray}
where $Z^{0}_{N}$ is the partition function for ideal gas. Knowing
it, we can find a decomposition of free energy on cluster's radius
$d$ which plays a part correlation length
\begin{eqnarray}
F&=&F_{0}-kT\left[\frac{V}{\lambda^{3}}\xi_{q}
-N\ln\left(1+\frac{\xi_{q}}{\xi_{0}}\right)
+\frac{V}{\lambda^{3}}\xi^{2}
d^{2}\alpha^{2}\right]\nonumber\\
&=&F_{0}+kT\left[-\frac{\alpha^{2}\lambda^{3}N^{2}}{V}d^{2}+\frac{\alpha^{4}\lambda^{6}N^{3}}{V^2}d^{4}\right]\texttt{,}
\label{2.11}
\end{eqnarray}
where $F_{0}$ is free energy for ideal gas. We can see that the
$d^{2}$ term is always negative
$-\frac{\alpha^{2}\lambda^{3}N^{2}}{V}<0$, that is cluster exists
in the system with Coulomb attraction at any thermodynamical
conditions. Minimizing (\ref{2.11}) by the size of the cluster
\begin{equation}
  \frac{\partial F}{\partial d}=-kT\frac{d\alpha^{2}\lambda^{3}2N^{2}}{V}
  \left[1-4\frac{d^{2}\alpha^{2}\lambda^{3}N}{V}\right]=0\texttt{,}
  \label{2.12}
\end{equation}
we obtain the optimum radius of the cluster,
\begin{equation}\label{2.13}
    d^{2}_{0}=\frac{V}{4Nr^{3}_{q}}
\end{equation}
or, for the dimension value $D=d\cdot r_{q}$,
\begin{equation}\label{2.14}
    D^{2}_{0}=\frac{1}{4}\frac{kT}{2\pi q^{2}}\frac{V}{N}\texttt{.}
\end{equation}

This expression means that equilibrium size of the cluster is
defined by balance of the two forces. First of them is attraction
Coulomb energy, which aspires to compress the gas. It is
represented by the multiplier $q^{2}$ (in sense of interaction
constant). The decrease of the cluster's size with the increase of
average density in the system $N/V$ is connected with a closer
packing of the particles in the cluster on account of the increase
of the energy of interaction. The second of them is the thermal
energy, which creates positive pressure resisting to the
compression. It is represented by the multiplier $kT$.With
increase of temperature the size of the cluster $d$ is rising,
however the central density is decreasing.When $T\rightarrow
\infty$ the radius of the cluster $d\rightarrow\infty$ and the
central density $\sigma(0)\rightarrow 1$. Such behavior the value
$d$ tells us that $d$ plays a part of correlation length which in
the point of phase transition (in our case is $T\rightarrow
\infty$) in infinitely large. Such situation is realized due to
the long-range type ($\sim 1/r$) of Coulomb interaction.

On other hand Eq.(\ref{2.14}) may be understood as such such
distance in the system on which the essential deflection from the
average fixed density $N/V$ is observed. If we assuming that
$q^{2}=Gm^{2}$, where $m$ is particle mass and $G$ is
gravitational constant, this expression may be understood as Jeans
length \cite{jea}.

Let's consider the dynamics of cluster's formation. For it we'll
use the equation of motion as
\begin{equation}
\frac{\partial d}{\partial t}=-\chi\frac{\partial F}{\partial d},
\label{2.14a}
\end{equation}
where $\chi$ is a some constant. Applying (\ref{2.11}) and
(\ref{2.13}) we have:
\begin{equation}\label{2.14b}
\dot{d}+\eta d^{3}-\eta d^{2}_{0}d=0,
\end{equation}
where we marked in $\eta\equiv\frac{\chi NkT}{2d^{4}_{0}}$. The
solution of Eq.(\ref{2.14b}) is
\begin{equation}\label{2.14c}
d^{2}=\frac{d^{2}_{0}}{1+C\exp\left(-2\eta d^{2}_{0}t\right)},
\end{equation}
where $C$ is integration constant. Let's consider the next initial
states.
\begin{itemize}
    \item Let the radius of a spatial
    inhomogeneity was larger than the equilibrium size $d(t)>d_{0}$ in some moment of time
    $t>0$. Then the constant $-1<C<0$, that means the cluster's increase from spatially homogeneous
distribution (where $d\rightarrow\infty$) to the equilibrium size
$d_{0}$.
    \item Let the radius of a spatial
    inhomogeneity was smaller than the equilibrium size
    $d(t)<d_{0}$. Then the constant $C>0$, that
that means the cluster's evaporation from initial distribution
with central density which is larger then equilibrium density at
the given temperature.
\end{itemize}
We can see that size of the cluster approaches to the equilibrium
size (\ref{2.13}) asymptotically $d=d_{0}[1+C\exp (-2\eta
d^{2}_{0}t)]$ when deflections are small. The term $1/2\eta
d^{2}_{0}$ can be understood as time of relaxation. If some
fluctuation of density has appeared in the system, then it brings
about the appearance of the potential's gradient. In its turn, it
brings about the spatial inhomogeneity - cluster with the size
approaching to the equilibrium value asymptotically. In others
words, cluster is equilibrium structures if we are assuming that
the average energy of interaction of two particles is less than
the average kinetic energy $\sim kT$ of a particle
$\frac{r^{3}_{q}N}{V}\ll 1$, when the term
$\frac{2}{r}\frac{\partial\varphi}{\partial r}$ in Eq.(\ref{2.6})
can be omitted, as it will be in the next subsection.

Since the number of particles in the system $N\rightarrow\infty$,
but the number of the particles in the cluster is finite, then
this means that our system disintegrates to the infinity multitude
of clusters of the size $D_{0}$ each. After that the process of
cluster formation repeats again, where the early formed clusters
plays a part of particles. In others words, the free anergy of
such system hasn't absolute minimum and each state of the system
is analogous to false vacuum in theory of field \cite{col}.

\subsubsection{Single-particle motion.}
\label{motion}

Let's consider single-particle motion in order to ground our
assumptions in the previous section for calculation of cluster's
size. For evaluation of cluster's size we can make use the another
method without calculation of the free energy. Let's we assume
that the average energy of Coulomb interaction of two particles is
less than the average kinetic energy $\sim kT$ of a particle, when
$\frac{r^{3}_{q}N}{V}\ll 1$ that has been named by "local ideality
approach" \cite{veg2}. This must lead to the two consequences:
\begin{itemize}
  \item the activity must be as
\begin{equation}\label{2.15}
  \xi=\xi_{0}\left(1+O\left[\frac{r^{3}_{q}}{V/N}\right]\right),
\end{equation}
where $\xi_{0}=\frac{\lambda^{3}N}{V}$ is activity for ideal gas.
That is the system's activity differences from activity for ideal
gas little.
  \item The center density (Fig.\ref{fig1}) differences from the
  average density little: $\rho(0)\simeq mN/V$. Then we can
  think that $\Delta^{2}\simeq \frac{N}{V}r_{q}^{3}$ and we can
  evaluate the term $\frac{2}{r}\frac{\partial\varphi}{\partial
  r}$ or $\frac{2}{r}\frac{\partial\sigma}{\partial r}$ in
  Eqs.(\ref{2.6},\ref{2.6b}) with help the solution
  (\ref{2.7}):\par
  $\frac{\max\left|\frac{2}{r}\frac{\partial\sigma}{\partial r}\right|}{\frac{\partial^{2}\sigma}{\partial
r^{2}}-\frac{1}{\sigma}\frac{\partial\sigma}{\partial
r}+\xi\frac{r_{q}^{3}}{\lambda^{3}}\sigma^{3}}\sim\frac{r^{3}_{q}N}{V}\ll
1$. That is this term can be omitted in the local ideality
approach.
\end{itemize}

\begin{figure}
\centering
    \includegraphics[width=8cm]{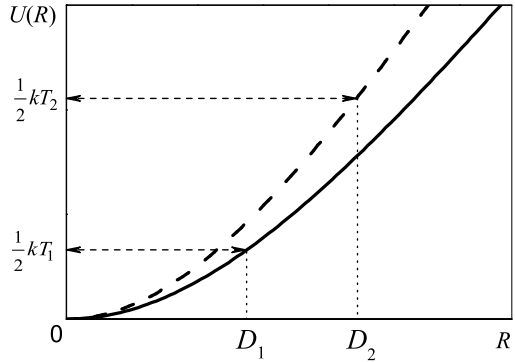}
\caption{The potential energy $U(r)$ of a particle in the cluster
(schematically). The solid line corresponds to a lower
           temperature $T_{1}$, the dash line corresponds to a higher temperature $T_{2}$.
           The particle does finite motion in the potential well created
by the all particles in the system. Then the radius of the cluster
$d$ we determine as distance from the cluster's center to the
point of stop of a particle with average kinetic energy
$\frac{1}{2}kT$.}\label{fig2}
\end{figure}

After that let's calculate potential energy of a particle $U(r)$
in the cluster using Boltzmann distribution
\begin{equation}\label{2.16}
  \sigma(r)=\sigma(0)\exp\left[-\frac{U(r)}{kT}\right].
\end{equation}
Then using (\ref{2.7}) we have that
\begin{eqnarray}\label{2.17}
U(r)=&&-kT\ln\left[\frac{\sigma(r)}{\sigma(0)}\right]=
2kT\ln\left[\cosh\left(\Delta r\right)\right].
\end{eqnarray}
The field $U(r)$ is self-coordinated field of the rest particles
in the system (in the cluster). Using the assumption (\ref{2.15})
about interactional and thermal energies and considering small
distances from the center of the cluster as $r/r_{q}\leq1$, we can
write that
\begin{equation}\label{2.18}
  U(r)\approx
  kT\Delta^{2}r^{2}\approx
  \frac{kTNr_{q}^{3}}{V}r^{2}.
\end{equation}
We can see that a particle in the cluster is in potential well
created by all cluster's particles. Then cluster's radius $d$ can
be understood as such distant from center of the cluster that
\begin{equation}\label{2.19}
  U(d)=\frac{1}{2}kT,
\end{equation}
where, the multiplier 1/2 corresponds to radial degree of freedom.
This means that $d$ is distance from the cluster's center to the
point of stop of a particle with average kinetic energy
$\frac{1}{2}kT$ (Fig.\ref{fig2}). That is radius of the cluster we
calculate by self-coordinated manner.

Then solving the equation (\ref{2.19}) we has the equilibrium
radius of the cluster as $d^{2}=\frac{V}{2r_{q}^{3}N}$, which
coincides with Eq.(\ref{2.13}) with exactness to constant
multiplier. Since two particles attracting by low $1/r$, they can
not fall on each other in Newton mechanics. Hence spatial
distribution is regular everywhere in the limits of the local
ideality approach, however it is not so in the total case
\cite{isp}. Such mechanism of cluster formation we shall call by
the thermal mechanism. It is necessary to notice that
single-particle motions in cluster is superdiffusive \cite{tor}.

\subsection{The cluster's size at all temperatures.}
\label{sec:alltemperatures}

In order to calculate radius of a cluster at any temperatures we
must solve Eq.(\ref{2.3}) and obtain spatial distribution function
of type (\ref{2.7}). We can't do this by cause mathematical
difficulties, however, if we suppose that gas is ideal locally,
that is $\frac{r_{q}^{3}N}{V}\ll 1$, and considered small
distances from the coordinates' center $R/r_{q}\leq 1$ where
$|\varphi|\ll 1$ (see Eq.(\ref{2.6a}) and Eq.(\ref{2.7})), then
our problem simplifies.

For example, let's consider Eq.(\ref{2.6}) and decompose
$e^{\varphi}$ in a row:
\begin{eqnarray}\label{2.20}
\frac{\partial^{2}\varphi}{\partial
r^{2}}+\xi_{0}\frac{2r_{q}^{3}}{\lambda^{3}}+O(\varphi)=0,
\end{eqnarray}
where we suppose that $\xi\approx\xi_{0}$. Its solution and
corresponding density is
\begin{eqnarray}
\varphi\approx\label{2.21}
-\xi_{0}\frac{2r_{q}^{3}}{\lambda^{3}}r^{2}\Rightarrow\rho=\frac{\xi}{\lambda^{3}}\exp\left(-\frac{\xi_{0}r_{q}^{3}}{\lambda^{3}}r^{2}\right).
\end{eqnarray}
Then we can obtain potential energy of a particle in the cluster
as
\begin{equation}\label{2.22}
  U(r)=-kT\ln\left[\exp\left(-\frac{\xi_{0}r_{q}^{3}}{\lambda^{3}}r^{2}\right)\right]=kT\frac{\xi_{0}r_{q}^{3}}{\lambda^{3}}r^{2},
\end{equation}
which has be obtained early (\ref{2.18}). Then, using the
condition $U(d)=\frac{1}{2}kT$ and expression for the Boltzmann
activity $\xi_{0}=\frac{\lambda^{3}N}{V}$ we obtain result
\begin{equation}\label{2.23}
  kT\frac{\xi_{0}r_{q}^{3}}{\lambda^{3}}d^{2}=\frac{1}{2}kT\Rightarrow d^{2}=\frac{V}{2r_{q}^{3}N},
\end{equation}
which coincides with Eq.(\ref{2.13}).

Now let's to calculate the radius of cluster for any temperatures
using Eq.(\ref{2.3}) and decomposing one in a row as
\begin{eqnarray}\label{2.24}
\frac{\partial^{2}\varphi}{\partial
r^{2}}+\frac{2r_{q}^{3}}{\lambda^{3}}\left(\frac{\partial
f_{5/2}\left(\xi_{0}
e^{\varphi}\right)}{\partial\varphi}\right)_{\varphi=0}+O(\varphi)=0.
\end{eqnarray}

Analogously, the size of the spatial inhomogeneity is determined
by the term at zero degree of the field $\varphi$. However, the
calculations with spatial distribution function (\ref{2.2a}) is
somewhat problematical because the underintegral expression can
not be integrated in elementary function. That's why we have to
find another method. If we suppose that gas is ideal locally, that
we have to suppose that motion of particles is quasiclassical
\cite{lan}. Than we propose to use the particle's energy in
Boltzmann view
\begin{equation}\label{2.25}
  U(r)=-kT\ln\left[\frac{\rho(r)}{\rho(0)}\right]=-kT\ln
  e^{\varphi (r)}=-kT\varphi (r),
\end{equation}
where the distribution (\ref{2.6a}) has been used and the field
$\varphi$ is solution of the Eq.(\ref{2.24}), \underline{however}
the radius of the cluster $d$ we determine as
\begin{equation}\label{2.26}
U(d)=\frac{1}{2}kT+\frac{3}{5}\varepsilon_{F},
\end{equation}
where $\varepsilon_{F}$ is Fermi energy which is determined by the
fixed average density of the system (\ref{A8}). The term
$\frac{1}{2}kT$ is a particle's kinetic energy caused by thermal
motion and $\frac{3}{5}\varepsilon_{F}$ is average particle's
kinetic energy in degenerated Fermi gas (because all particles can
not be in the level $\varepsilon=0$ by force of Pauli principle).
The term $\frac{3}{5}\varepsilon_{F}$ can be understood as
repulsive ''statistical potential'' \cite{hau} too. Then the
assumption about local ideality of gas is written as
$\frac{q^{2}}{\sqrt[3]{V/N}}\ll kT+\varepsilon_{F}$ - the average
energy of interaction of two particles is less than the average
kinetic energy of a particle.

Using the decomposition (\ref{2.24}), equation (\ref{2.26}) and
equation (\ref{A2}) for determination of the activity $\xi_{0}$ we
can obtain expression for radius of the cluster for Coulomb
interacting fermi-gas:
\begin{equation}\label{2.27}
\left[\begin{array}{c}
  kT\frac{2r_{q}^{3}}{\lambda^{3}}\left(\frac{\partial
f_{5/2}\left(\xi_{0}
e^{\varphi}\right)}{\partial\varphi}\right)_{\varphi=0}d^{2}=\frac{1}{2}kT+\frac{3}{5}\varepsilon_{F} \\
\\
  \frac{\lambda^{3}N}{V}=f_{3/2}(\xi_{0})
\end{array}\right].
\end{equation}

Let's consider some limit cases. In the limit of high temperatures
we supposed that $kT\gg\varepsilon_{F}$. Then calculation the
first quantum correction only in the decomposition (\ref{A2}) we
can obtain that
\begin{equation}\label{2.28a}
  d^{2}=\frac{V}{2r_{q}^{3}N}+\frac{1}{2^{5/2}}\frac{\lambda^{3}}{r_{q}^{3}},
\end{equation}
where $\lambda/r_{q}\ll 1$. For the dimension value:
\begin{equation}\label{2.28b}
  D^{2}=\frac{kTV}{4\pi
  q^{2}N}+\frac{1}{2^{7/2}}\frac{\hbar^{3}}{\pi (2\pi
  m)^{3/2}(kT)^{3/2}}.
\end{equation}
The sense of the first addendum has been explained in the section
(\ref{sec:BoltzmannLimit}). The sense of the first quantum
correction is explained below.

It is know that so-called ''statistical potential'' of particles'
interaction exists in quantum gases \cite{hau}. It is repulsion
for fermions and it is attraction for bosons. This phenomenon is
connected with symmetry of particles' wave function. Hence the
equilibrium size is determined by three energies - of Coulomb
interaction (the interaction constant is $r_{q}$ or $q^{2}$),
thermal $kT$ and above-mentioned ''statistical potential''
($\lambda$ or $\hbar$ plays a part of interaction constant). That
is, in case of Fermi statistics, the quantum energy resists to the
compression under action of Coulomb force as the small addendum to
thermal energy for high temperatures. This explains the sign
''+''for the small correction to the cluster's radius. For case
Bose statistic we can analogous expression but the sign "+"
replaced by the sign "$-$".

In the limit of low temperatures $kT\ll\varepsilon_{F}$  we obtain
that
\begin{equation}\label{2.29}
  D^{2}=0.4\frac{\hbar^{2}}{q^{2}m}\left(\frac{V}{N}\right)^{1/3}
  \left[1+\frac{5}{6}\frac{kT}{\varepsilon_{F}}\right].
\end{equation}
We can see that the equilibrium size is determined by two energies
- Coulomb potential and the repulsion statistical potential (the
constant of interaction is $\hbar$). That is, the aforesaid
quantum energy resists to the Colombian compression. The small
correction $\frac{kT}{\varepsilon_{F}}$ means that above-mentioned
thermal mechanism of cluster formation (as small addendum to Fermi
statistical repulsion) including if $T\neq 0$.

Expression for radius of a cluster at low temperatures can be
obtained by the next formal means. It is not difficult to notice,
that the equilibrium radius of a cluster is determined by the
multiplier $\frac{2r_{m}}{\lambda ^{3}}$ in the equation
(\ref{2.6}) after the term $e^{\varphi }$ at $\xi\approx\xi_{0}$.
In the case $T\rightarrow 0$ we can suppose that the radius is
determined by above-mentioned multiplier, however because
temperature is zero, hence $kT$ can not enter into the searching
formula. Then it is necessary to multiply this coefficient to the
value $\frac{kT}{\varepsilon _{F}}$ which cancellates temperature.
Then the cluster's radius is determined as
$\frac{1}{D^{2}}\propto\frac{r_{q}}{\lambda ^{3}}\left(
\frac{kT}{\varepsilon
_{F}}\right)=\frac{\hbar^{2}}{q^{2}m}\left(\frac{V}{N}\right)^{1/3}$,
and we have the formula (\ref{2.29}).

\section{Formation of structures in interacting particle's system with Coulomb repulsion}
\label{sec:repulsion}
\subsection{Statistical approach}

Let $N_{+}$ particles repulse by Coulomb low with the interaction
constant $q^{2}$. In order to ensure stability of such system, we
suppose that these particles is dipped in environment of contrary
sign (with charge of $-q$), so that the condition of
neutralization is executed: $N_{+}=N_{-}=N$, but a current created
by $N_{+}$ particles only. In other worlds, we consider such
plasma where constant of interaction is $q^{2}$ and charge's
carriers of one from the signs is immovable. Let's find spatial
distribution function for such system, where
$N\rightarrow\infty,V\rightarrow\infty$ and $N/V$ is fixed.

Let a particle with charge $q$ is situated in the beginning of
coordinates. It polarizes the gas of particle with the same charge
(the environment with contrary sign is immovable). For repulsion
particles the action (\ref{1.3}) has view:
\begin{equation}\label{3.2}
  S=\int^{V}_{0}\left[\frac{(\nabla\psi)^{2}}{4r_{q}}-\frac{1}{\lambda^{3}}f_{5/2}(\xi\cos\psi)\right]dV+N\ln\xi,
\end{equation}
and corresponding Lagrange equation is
\begin{equation}\label{3.3}
\frac{d^{2}\psi}{dR^{2}}+\frac{2r_{q}}{\lambda^{3}}\frac{\partial
f_{5/2}\left(\xi\cos\psi\right)}{\partial\psi}=0.
\end{equation}
The term $\frac{2}{r}\frac{\partial\psi}{\partial r}$ can be
omitted as in the case of attracting particles. In Boltzmann limit
$\xi\rightarrow 0$ we can write the action and spatial
distribution function as
\begin{eqnarray}
S=\int^{V}_{0}\left[\frac{(\nabla\psi)^{2}}{4r_{q}}-\frac{\xi}{\lambda^{3}}\cos\psi\right]dV+N\ln\xi,\\
\label{3.4} \rho(R)=\frac{\xi}{\lambda^{3}}\cos\psi\label{3.5},
\end{eqnarray}
\begin{figure}
\includegraphics[width=8cm]{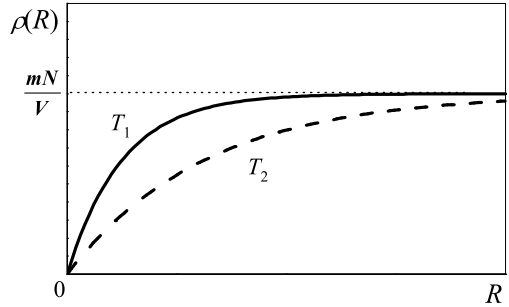}
\caption{The density $\rho(R)$ of polaron at different
        temperatures (schematically). The solid line corresponds to a
        lower temperature ($T_{1}$), the dash line corresponds to a higher
        temperature ($T_{2}$). The dot line $\rho=N/V$ corresponds to
        average fixed density.}\label{fig3}
\end{figure}
with corresponding Lagrange equation
\begin{equation}\label{3.6}
\frac{d^{2}\psi}{dR^{2}}-\frac{2r_{q}}{\lambda^{3}}\xi\sin\psi=0.
\end{equation}

We assume that average energy of interaction of two particles is
much less than their average kinetic energy, that is
$\xi\approx\xi_{0}=\frac{\lambda^{3}N}{V}$, and we find solution
in the area $\psi(R)\in[0,\pi/2]$ where $R\in[0,+\infty]$ with
asymptotics: $\psi (0)=\pi /2$ and $\psi (+\infty)=0$. The first
of the asymptotics corresponds to absence of particles in the
beginning of coordinates, because in order to penetrate there a
particle must has infinite kinetic energy. Second of them means
that the concentration equals to average concentration on large
distances from the center $\rho(R\rightarrow\infty)=N/V$. In other
words, the spatial distribution function must be pit - such
formation called as polaron usually, unlike the system with
attraction where spatial distribution is hump - cluster. In other
words polaron can be called by anticluster. Corresponding solution
of Eq.(\ref{3.6}) is
\begin{equation}\label{3.7}
 \tan\frac{\psi}{4}=\tan\frac{\pi}{8}
 \exp\left(-\sqrt{\frac{2\xi_{0}r_{q}}{\lambda^{3}}}R\right).
\end{equation}

Then spatial distribution function (Fig.\ref{fig3}) is written as
\begin{equation}\label{3.7a}
  \rho(R)=\frac{N}{V}\cos\left[4\arctan\left(
  \exp\left(-\sqrt{\frac{2\xi_{0}r_{q}}{\lambda^{3}}}R\right)\tan\frac{\pi}{8}\right)\right].
\end{equation}
This distribution is determined by balance of two energies:
Coulomb energy which prevents from penetration of a particle in
the center of coordinates, and the thermal energy (average kinetic
particle's energy). Then radius of the polaron we can find as
distance from its center to the point of stop: $U(D_{p})\sim kT$.
The function $U(r)$ we can find from Boltzmann distribution
\begin{equation}\label{3.8}
  \rho(R)=\rho(R\rightarrow\infty)\exp\left(-\frac{U(R)}{kT}\right).
\end{equation}
It is using (\ref{3.7}) we have
\begin{equation}\label{3.9}
  U=-kT\ln\left[\cos\left[4\arctan\left(
  \exp\left(-\sqrt{2\xi_{0}\frac{r_{q}}{\lambda^{3}}}R\right)\tan\frac{\pi}{8}\right)\right]
  \right].
\end{equation}
In order to find radius of the polaron let's apply the equation
$U(D_{p})=\frac{1}{2}kT$ and we obtain that
\begin{equation}\label{3.10}
  D^{2}_{p}\propto\frac{kTV}{4\pi q^{2}N}.
\end{equation}
This expression for polaron's radius coincides with cluster's
radius (\ref{2.14}) with exactness to the constant multiplier.
This value we can understand as the size on which the
neutralization of gas is violated.

\begin{figure}
  \includegraphics[width=9cm]{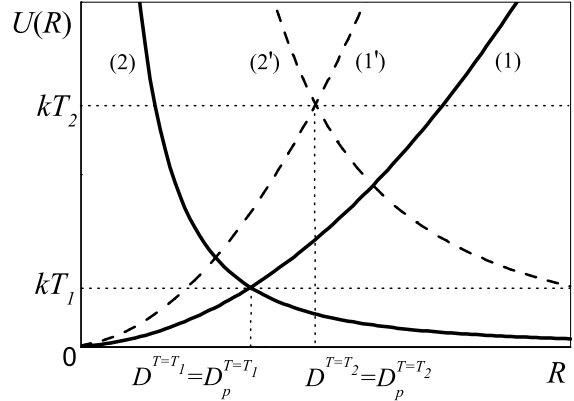}\\
  \caption{The potential energy of a gravitating particle in a cluster - the curves \textbf{(1)} and \textbf{(1')}
and a particle in a polaron - the curves \textbf{(2)} and
\textbf{(2')} at temperatures $T_{1}$ and $T_{2}$ accordingly. In
the Boltzmann limit the radii of the cluster $D$ and the polaron
$D_{p}$ is determined as the points of stop of a particle with the
average kinetic energy $kT$, which explains these
equality.}\label{Fig4}
\end{figure}

The coincidences sizes $D$ and $D_{p}$ we can explain by the next
way. We have seen that the spatial distributions in the both cases
is result of balance of two energies only: Coulomb type and
thermal energy. The radii of cluster and polaron are determined by
the points of stop of a particle in the field of the rest
particles. That is mechanism of formation of such structures is
one and the same - a balance of they two energies. They are
differed by roles only: in the gas of attracting particles they
make spatial distribution of view "hump" and in the case of
repulsing particles - "pit" Fig.(\ref{Fig4}).

In the degenerated case situation is analogous, only the Fermi
energy plaices a part of the thermal energy, that is the equation
for the point of stop is $U(r)\sim\varepsilon_{F}$. Let's use the
method described in subsection (\ref{sec:alltemperatures}) in
order to find radius of the polaron. This size is determined by
the multiplier $\frac{2r_{q}}{\lambda ^{3}}$ in Eq.(\ref{3.3}),
but we must get rid of temperature in the final expression. For
this let's multiply $2r_{q}/\lambda ^{3}$ to the dimensionless
value $kT/\varepsilon _{F}$, where $\varepsilon _{F}$ is Fermi
energy (\ref{A8}) determined by the average density of the system
$N/V$. Then the polaron's radius is wrote as
\begin{equation}\label{3.11}
\frac{1}{D_{p}^{2}}\propto \frac{r_{q}}{\lambda ^{3}}\left( \frac{kT}{%
\varepsilon _{F}}\right) \propto \frac{q^{2}m}{\hbar ^{2}}\left( \frac{N}{V}%
\right)^{1/3}.
\end{equation}
This expression coincides with the expression for radius of
cluster (\ref{2.29}) without the correction for temperature in the
system of attracting particles. Such coincidence is explained by
the same as in classical case, but a part of the average kinetic
energy $kT$ plays Fermi energy, which we can understand as
above-mentioned "statistical potential" too.

\subsection{Hydrodynamical approach}
\label{sec:HidrodynamicalApproach}

In order to explain the coincidence of the radii of cluster $D$
and polaron $D_{p}$ we can use hydrodynamical approach \cite{jea}
which lets to look at some general properties of system with
Coulomb type interaction under other point of view. Let's consider
the system of particle which is described be the persistent
equation, Euler equation and Poisson equation (''+'' for
gravitating particles and ''$-$'' for interaction like electrical
charges):
\begin{eqnarray}\label{3.11}
    \frac{\partial \rho }{\partial t}+\nabla \cdot \left( \rho
    \overrightarrow{v}\right) \nonumber\\
    \rho \frac{\partial \overrightarrow{v}}{\partial t}+\rho \left(
\overrightarrow{v}\cdot \nabla \right) \overrightarrow{v}=-\nabla
P+\rho \nabla \phi\\
\Delta \phi =\pm 4\pi G\rho \text{.}\nonumber
\end{eqnarray}
We suppose that perturbations is adiabatical:
$P_{1}=\frac{\partial P}{\partial \rho }\rho _{1}=c^{2}\rho _{1}$,
where $c=\sqrt{kT/m}$ - the speed of sound, $P_{1}\ll P$ - the
small change of pressure. Then the linearized equation of motion
for fluctuation $\rho$ is
\begin{equation}\label{3.12}
\left( \Delta -\frac{1}{c^{2}}\frac{\partial ^{2}}{\partial t^{2}}%
\right) \rho _{1}\pm k_{J}^{2}\rho _{1}=0\text{,}
\end{equation}
where $k_{J}=\frac{\sqrt{4\pi G\rho _{0}}}{c}$ - Jeans wave
number.

Let's described localized oscillation process, when $k=0$. Then we
have next solution:
\begin{itemize}
    \item For the system with Coulomb repulsion the dynamics of increase of the  deflection
    has view  $\rho _{1}\propto \exp \left(
i\omega t\right)$, where $\omega =ck=\sqrt{\frac{4\pi Nm}{V}}$.
That is we have plasma oscillations.
    \item for self-gravitation system the dynamics of increase of the  deflection
    has view $\rho _{1}\propto \exp \left( \omega t\right)$ or  $\rho
_{1}\propto\exp\left( -\omega t\right)$,
    where $\omega =ck=\sqrt{\frac{4\pi Nm}{V}}$.That means exponential
increase of the  deflection or relaxation to the equilibrium state
- cluster.
\end{itemize}
We can see, that the both process is characterized by one and the
same time scale $\tau \sim \frac{1}{\omega }$/ Then we can
suppose, that the correspond spatial scale is distant, which a
particle cross in the process of its thermal motion at the
time$1/\omega$:
\begin{equation}\label{3.13}
D\propto \frac{<\upsilon >}{\omega }=\sqrt{\frac{kT}{m}}/\omega =\sqrt{\frac{%
kTV}{4\pi Gm^{2}N}}\text{.}
\end{equation}
This expression coincides with radiuses $D$ and $D_{p}$ for
gravitating particles and repulsing particles accordingly. That is
both aforesaid systems is characterized by one and the same sizes
of spatial inhomogeneity.

As it has been said above, the partition function of
self-gravitational system divergences, however the states exist
where the equilibrium spatial distribution function is solution of
Eq.(\ref{2.6}). This means that our system disintegrates
(collapses) to infinity multitude of clusters of size $D_{0}$
each, since the number of particles in the system
$N\rightarrow\infty$, but the number of the particles in the
cluster is terminal. This multitude is self-gravitating system too
and it may be collapse. Then the process repeats again. In others
words, the free anergy of the such system hasn't absolute minimum
and each state of the self-gravitating system is analogous to
false vacuum in theory of field \cite{col}. To evaluate of times
of relaxation we can using Eq.(\ref{3.13}). So with each act of
collapse, we must substitute $m\rightarrow nm$ and $N\rightarrow
N/n$, where $n$ is average number of particles in a cluster. Then
the time scale $1/\omega$ is constant at each process of cluster
formation (collapse).

\section{Conclusion.}

In this article, based on statistical approach
\cite{zhu,lev,bel,bel1}, we demonstrated a nonperturbative
calculation of the partition function, we solved the system of
particles with Boltzmann and Fermi statistics with long-range
interaction of Coulomb type. All results have been obtained at the
condition that average energy of interaction of two particles is
much less than their average kinetic energy $kT$ or Fermi energy
$\varepsilon_{F}$: $\frac{q^{2}}{\sqrt[3]{V/N}}\ll kT$ or
$\frac{q^{2}}{\sqrt[3]{V/N}}\ll\varepsilon_{F}$ - the "local
ideality" approach \cite{veg2}.

The long-range attraction of particles results to the formation of
the cluster of finite size, as the initial homogeneous state is
instable in the limit $N\rightarrow\infty$, $V\rightarrow\infty$
with fixed $N/V$. The equation for spatial distribution function
has be obtained. However, it is simplified in the next limit cases
only .

In case of large temperature (when $\lambda\ll\sqrt[3]{\upsilon}$)
the radius of cluster is determined by balance of the two
energies: the energy of attraction interaction which aspires to
compress gas, and the thermal energy which creates a positive
pressure. This mechanism created equilibrium distribution in the
the system (\ref{2.7}), in particular, with rising of temperature
the cluster's size enlarges and its central density decreases
accordingly. In the case of degenerated system the radius of the
cluster is determined by balance of the two energy too: energy of
interaction which aspires to compress gas, and repulsive Fermi
statistical potential creating positive pressure. In the total
case the equilibrium cluster's size determined by three energy:
energy of interaction, thermal and repulsive "statistical
potential". The size of the cluster approaches to the equilibrium
size asymptotically in process of its formation. The state of the
system with spatially inhomogeneous distribution corresponding to
the cluster of the equilibrium size (\ref{2.28b} or \ref{2.29}) is
stable if we are assuming that "local ideality" approach is true.

The size of spatial inhomogeneity of two matter: gravitating gas
and repulsing particles (by Coulomb) is coincides at one and the
same interaction's constant in Boltzmann limit and degenerated
case. This coincidence caused by that, that the spatial
distributions in both cases is determined by the two energies
only: Coulomb type and thermal energy or repulsive Fermi
''statistical potential''. The mechanism of formation of the
cluster and the polaron is one and the same - a balance of they
two energies. This fact can be explained by hydrodynamical
approach - both matters have same time scale of structure's
formation, but in the first matter it determines time of
relaxation and in the second matter it determines of plasma
oscillation. This time scales determine spatial scale (the radii
of the cluster and the polaron) monotonously.

\appendix
\section{The effective free energies for ideal quantum gases.}

Let's obtain some expressions used in this paper for ideal Fermi
and Bose gas. Thus we'll demonstrate the correctness and greater
rationality of our approach as compared with the traditional
method \cite{hau}.

In the case of ideal gas $\varphi=\psi=0$. Then the effective free
energy (\ref{1.3}) for the system is

\begin{eqnarray}
S=&&-\frac{1}{\omega}\int dV\int\ 4\pi
p^{2}dp\ln\left(1+\xi\exp\left(-\beta\varepsilon_{p}\right)\right)+N\ln\xi\nonumber\\
&&=-\frac{V}{\lambda^{3}}f_{5/2}\left(\xi\right)+N\ln\xi
\label{A1},
\end{eqnarray}
where
\begin{eqnarray}\label{A1a}
  f_{5/2}\left(\xi\right)&&=\frac{4}{\sqrt{\pi}}\int^{\infty}_{0}dxx^{2}\ln\left(1+\xi
e^{-x^{2}}\right)\nonumber\\
&&=\sum^{\infty}_{l=1}(-1)^{l+1}\frac{\xi^{l}}{l^{5/2}}
\end{eqnarray}
is special Fermi function \cite{hau,isi};
$\varepsilon_{p}=p^{2}/2m$ is kinetic energy of a particle;
$\lambda=\sqrt{\frac{\beta\hbar^{2}}{2\pi m}}$ is wave's thermal
length of a particle. Then the equation for saddle point is as
follows:
\begin{equation}
\frac{1}{v}=\frac{1}{\lambda^{3}}f_{3/2}\left(\xi\right)
\label{A2},
\end{equation}
where $f_{3/2}\left(\xi\right)=\xi\frac{\partial
f_{5/2}\left(\xi\right)}{\partial\xi}$, $v\equiv\frac{V}{N}$.

The partition function (\ref{1.2}) for this case:
\begin{equation}
    Z_{N}=\exp\left[\frac{V}{\lambda^{3}}f_{5/2}\left(\xi\right)-N\ln\xi\right],
\label{A3}
\end{equation}
where $\xi$ is determined from the equation (\ref{A2}). We can
find any thermodynamical functions knowing the partition function.
Let's consider the case $\xi\rightarrow 0$ corresponding to high
temperature (the Boltzmann limit), then Eqs. (\ref{A2},\ref{A3})
are reduced to
\begin{eqnarray}\label{A6}
  Z_{N}=\exp\left[\frac{V}{\lambda^{3}}\xi-N\ln\xi\right]
\approx\frac{V^{N}}{N!}\left(\frac{mkT}{2\pi\hbar^{2}}\right)^{\frac{3}{2}N}\label{A6}\\
\nonumber\\
\frac{1}{v}=\frac{\xi}{\lambda^{3}}\label{A7}
\end{eqnarray}

Let's consider the case $T\rightarrow 0$ corresponding to
degenerated Fermi gas. Then the special Fermi functions are
\cite{hau}
\begin{eqnarray}\label{A8}
f_{3/2}=\frac{4}{3\sqrt{\pi}}\left[(\ln\xi)^{3/2}+\frac{\pi^{2}}{8}(\ln\xi)^{-1/2}\right]+O(\xi^{-1})\nonumber\\
f_{5/2}=\frac{4}{3\sqrt{\pi}}\left[\frac{2}{5}(\ln\xi)^{5/2}+\frac{\pi^{2}}{4}(\ln\xi)^{1/2}\right]+O(\xi^{-1}).
\end{eqnarray}
In this case, the expression (\ref{A2}) is reduced to
\begin{equation}\label{A9}
  \frac{\lambda^{3}}{\upsilon}\approx\frac{4}{3\sqrt{\pi}}(\ln\xi)^{3/2}\Rightarrow\xi\approx e^{\beta
  \varepsilon_{F}},
\end{equation}
where
\begin{equation}\label{A10}
  \varepsilon_{F}\equiv\frac{\hbar^{2}}{2m}\left(\frac{6\pi^{2}}{\upsilon}\right)^{\frac{2}{3}}
\end{equation}
is Fermi energy.

We can find analogical expressions for Bose gas. It is necessary
to proceed from the effective free energy as
\begin{eqnarray}
S=&&\frac{1}{\omega}\int dV\int\ 4\pi
p^{2}dp\ln\left(1-\xi\exp\left(-\beta\varepsilon_{p}\right)\right)+N\ln\xi\nonumber\\
&&=-\frac{V}{\lambda^{3}}g_{5/2}\left(\xi\right)+\ln\left(1-\xi\right)+N\ln\xi,
\label{A11}
\end{eqnarray}
where
$g_{5/2}\left(\xi\right)=-\frac{4}{\sqrt{\pi}}\int^{\infty}_{0}dxx^{2}\ln\left(1-\xi
e^{-x^{2}}\right)=\sum^{\infty}_{l=1}\frac{\xi^{l}}{l^{5/2}}$ -
special Bose function \cite{hau,isi}. It is clear that activity
always $\xi<1$ unlike Fermi system; $\ln\left(1-\xi\right)$ is
effective free energy for condensed phase (the addendum with
$\textbf{p}=0$ is as important as the rest of the sum when
$\xi\rightarrow 1$).

Let's find the internal energy $U$ of the system proceeding from
the arranged analogy between apparatus of thermodynamics in our
representation
and the theory of field. In order to do it let's use the correlation $%
-H=\partial S_{mech}/\partial t$ and determine conformities $%
H\longleftrightarrow U,S_{mech}\longleftrightarrow S_{term},
t\longleftrightarrow 1/kT $, where $H$ is Hamilton's function of
the system,
$S_{mech}$ and $S_{term}$ are the action for mechanics and effective free energy for thermodynamics (%
\ref{1.3}) systems correspondingly, $t$ is time and $1/kT$ is
reverse temperature. Then, with the help of the expression
(\ref{2.1}), we have
\begin{equation}
U=-\frac{\partial S}{\partial\left( 1/kT\right)}=\frac{3}{2}\frac{VkT}{%
\lambda^{3}}f_{5/2}\left(\xi\right)=\frac{3}{2}PV  \label{A12}
\end{equation}

Thus the expressions for thermodynamical function of Fermi and
Bose gases obtained by our method coincide with ones obtained by
usual way \cite{hau,isi}, and it confirms the correctness of the
proposed approach.

\end{document}